\documentstyle[11pt]{article}

\begin{document}

\title{Contextual logic for quantum systems}
\author{{\sc Graciela Domenech} \thanks{%
Fellow of the Consejo Nacional de Investigaciones Cient\'{\i}ficas
y T\'ecnicas (CONICET) }\ $^{1}$ \ and {\sc Hector Freytes}
$^{2}$\\
}

\maketitle

\begin{center}
\begin{small}
1. Instituto de Astronom\'{\i}a y F\'{\i}sica del Espacio (IAFE)\\
Casilla de Correo 67, Sucursal 28, 1428 Buenos Aires, Argentina\\
2. Escuela de Filosof\'{\i}a - Universidad Nacional de Rosario,\\
Entre R\'{\i}os 758, 2000, Rosario, Argentina
\end{small}
\end{center}

\bibliographystyle{plain}

\vspace{1cm}
\begin{abstract}

\noindent

In this work we build a quantum logic that allows us to refer to
physical magnitudes pertaining to different contexts from a fixed
one without the contradictions with quantum mechanics expressed in
no-go theorems. This logic arises from considering a sheaf over a
topological space associated to the Boolean sublattices of the
ortholattice of closed subspaces of the Hilbert space of the
physical system. Differently to standard quantum logics, the
contextual logic maintains a distributive lattice structure and a
good definition of implication as a residue of the conjunction.

\end{abstract}

\begin{small}

\centerline{\em PACS numbers: 03.65.Ta, 02.10.-v }

\end{small}

\bibliography{pom}

\newtheorem{theo}{Theorem}[section]

\newtheorem{definition}[theo]{Definition}

\newtheorem{lem}[theo]{Lemma}

\newtheorem{prop}[theo]{Proposition}

\newtheorem{coro}[theo]{Corollary}

\newtheorem{exam}[theo]{Example}

\newtheorem{rema}[theo]{Remark}{\hspace*{4mm}}

\newtheorem{example}[theo]{Example}

\newcommand{\proof}{\noindent {\em Proof:\/}{\hspace*{4mm}}}

\newcommand{\qed}{\hfill$\Box$}

\newcommand{\ninv}{\mathord{\sim}} 

\newpage

\section{Introduction}

Quantum mechanics has profound conceptual difficulties that may be
posed in several ways. Nonetheless almost every problem in the
relation between the mathematical formalism and what may be called
``our experience about the behavior of physical objects'' can be
encoded in the question about the possible meaning of the
proposition ``the physical magnitude {\it A} has a value and the
value is this or that real number''. Already from the first
formalizations this point was recognized. For example, P.A.M.
Dirac stated in his famous book: ``The expression that an
observable `has a particular value' for a particular state is
permissible in quantum mechanics in the special case when a
measurement of the observable is certain to lead to the particular
value, so that the state is an eigenstate of the observable. It
may easily be verified from the algebra that, with this restricted
meaning for an observable `having a value', if two observables
have values for a particular state, then for this state the sum of
the two observables (if the sum is an observable) has a value
equal to the sum of the values of the two observables separately
and the product of the two observables (if this product is an
observable) has a value equal to the product of the values of the
two observables separately.'' \cite{Dirac}. This last point is the
requirement of the functional compatibility condition (FUNC), to
which we will return later. As long as we limit ourselves to speak
about measuring results and avoid being concerned with what
happens to nature when she is not measured, quantum mechanics
carries out predictions with great accuracy. But if we naively try
to interpret eigenvalues as the possible or actual values of the
physical properties of a system, we are faced to all kind of no-go
theorems that preclude this possibility. Most remarkably is the
Kochen-Specker (KS) theorem that rules out the non-contextual
assignment of values to physical magnitudes \cite{ks}. Of course,
to restrict the valuation to a subset of observables -typically to
a complete set of commuting observables (CSCO) which constitutes a
$context$- and refer to values of physical variables only in the
sense allowed by the mathematical formalism, ensures no
contradiction. So, a widely accepted position is to abandon
seeking to describe what nature at the quantum level is and use
the theory as a mere instrument of prediction. But there are also
different proposals to investigate how to assign objective and
measurable properties to a physical entity, i.e. how far we can
refer to physical objects without contradiction with quantum
theory. This paper is framed in that search.

Our proposal is to construct a logic to enable not only a Boolean
valuation in each fixed context but also that, once chosen certain
set of projectors of the spectral decomposition of the operators
(correspondingly, closed subspaces of Hilbert space $\mathcal{H}$)
that admits a global Boolean valuation, to make it possible to
refer at least partially to projectors pertaining to other
contexts with the least arbitrariness.

Let us be here concerned with the simplest cases: pure states of
the system are represented by normalized vectors of $\mathcal{H}$
and dynamical variables {\it A} by bounded self-adjoint operators
${\bf A}$ with discrete spectra. The possible results of the
measurement of a (sharp) magnitude {\it A} are the eigenvalues
$a_{i}$ pertaining to the spectrum $\sigma$(${\bf A}$) of its
associated operator ${\bf A}$. To each of the eigenvalues $a_{i}$
corresponds a projector ${\bf P}_i$ and correspondingly a closed
subspace of $\mathcal{H}$. Every ${\bf A}$ admits a spectral
decomposition

\[
{\bf A}=\sum_{i} a_{i}{\bf P}_i
\]

\noindent where the equality is considered as a convergence in
norm. So observables may be discomposed to give an exhaustive and
exclusive partition of the possible alternatives for the results
of measurements. The probability to obtain one of them in an
experimental procedure is given by the Born rule.

Now let us suppose the state of the physical system is an
eigenvector of a non degenerate observable ${\bf A}$ (i.e. ${\bf
A}$ constitutes a CSCO) so we know the eigenvalues of all
projectors ${\bf P}_1, \ {\bf P}_2, \ ..., \ {\bf P}_n, \ ...$  of
${\bf A}$ for the system in this state. If any ${\bf P}_i$ lies in
the spectral decomposition of another observable ${\bf B}$, then
this ``part'' of {\it B} can be valued in a Boolean way. It is
important to realize that this allows to refer to observables
pertaining to a CSCO from another CSCO. In categorical terms, this
will be related to the possible local sections of a sheaf
satisfying a certain kind of compatibility with respect to fixed
contexts, to be exactly stated in what follows. From this formal
analysis in terms of sheaves, we intend to build the mentioned
logic, which we will call {\it contextual logic}. This {\it
contextual logic} will allow us to formalize to what extent we can
consider as objective properties of a physical system those
represented by projectors pertaining to different contexts without
facing no-go theorems. We will use a categorial frame to develop
this logic as has been the case during the last years, when
applications of category theory tools to logical questions in
standard quantum mechanics have begun to appear (for example Isham
and Butterfield \cite{isham1}, \cite{isham2}, \cite{isham3},
\cite{isham4}, also in consistent histories approach
\cite{isham1997}, in the interpretation of the Sasaki hook as an
adjunction \cite{smets2001} and, in general, in the
Geneve-Brussels approach \cite{aerts1}, \cite{aerts2}
\cite{coecke}).

In section 2 we introduce basic notions about lattice theory and
topics in categories. We devote section 3 to discuss the problem
of the valuations of physical magnitudes pertaining to different
CSCCs. In section 4 we face the same problem from the point of
view of sheaves relating it with the dual spectral presheaf
introduced in \cite{isham1}. In section 5 we develop the
contextual logic in a Kripke style and intuitionistic way.
Finally, in section 6 we outline our conclusions.

\section{Basic Notions}

We recall from \cite{BD,Bir} and \cite{goldblatt,MAKM} some
notions of the lattice theory and categories that will play an
important role in what follows.

First, let $(A, \leq)$ be a poset and $X\subseteq A$.  $X$ is {\it
decreasing set}  iff for all $x\in X$, if $a\leq x$  then $a\in
X$. For each $a\in A$ we define the {\it principal decreasing set}
associated to $a$ as $(a] = \{x\in A: x\leq a \}$. The set of all
decreasing sets in $A$ is denoted by $A^+$, and it is well known
that $(A^+, \subseteq)$ is a complete lattice, thus $\langle A,
A^+  \rangle $ is a topological space. We observe that if $G\in
A^+$ and $a\in G$ then $(a]\subseteq G$. Therefore $B = \{(a]:
a\in A \}$ is a base of the topology $A^+$ which we will refer to
as the {\it canonical base}. If $X \subseteq A$, we denote by
$\partial X$ the border of $X$, $C(X)$ the complement of $X$ and
$X^\circ$ the interior of $X$.

Let ${\cal A}$ be a category. We denote by $Ob({\cal A})$ the
class of objects and by $Ar({\cal A})$ the class of arrows. Given
an arrow $f:a\rightarrow b$ , $a$ is called domain of $f$ ($a =
dom(f)$) and $b$ is called codomain of $f$ ($cod(f)$). We denote
by $[a,b]_{\cal A}$ the class of all arrows $a\rightarrow b$ in
the category and by $1_A$ the identity arrow over the object $A$.
${\cal A}$ is said to be {\it small category} iff  $Ob({\cal A})$
is a set. A partially ordered set $(P, \leq)$ gives rise to a
category with the elements of $P$ as objects, and with precisely
one arrow $p\rightarrow q$ iff $p\leq q$. In this case, $(P,
\leq)$ is a small category. The category whose objects are sets
and arrows are functions with the usual composition is denoted by
$Ens$.

Let $I$ be a topological space. A {\it sheaf} over $I$ is a pair
$(A, p)$ where $A$ is a topological space and $p:A \rightarrow I$
is a local homeomorphism. This means that each $a\in A$ has an
open set $G_a$ in $A$ that is mapped homeomorphically by $p$ onto
$p(G_a) = \{p(x): x\in G_a\}$, and the latter is open in $I$. It
is clear that $p$ is continuous and open map. If $p:A \rightarrow
I$ is a sheaf over $I$, for each $i\in I$, the set $A_i = \{x\in
A: p(x)= i\}$ is called the {\it fiber} over $i$. Each fiber has
the discrete topology as subspace of $A$. {\it Local sections} of
the sheaf $p$ are continuous maps $\nu: U \rightarrow A$ defined
over open proper subsets $U$ of $I$ such that the following
diagram is commutative:

\begin{center}
\unitlength=1mm
\begin{picture}(60,20)(0,0)
\put(8,16){\vector(3,0){5}} \put(20,10){\vector(0,-2){5}}
\put(4,12){\vector(1,-1){10}}

\put(2,16){\makebox(0,0){$U$}} \put(20,16){\makebox(0,0){$A$}}
\put(20,0){\makebox(0,0){$U$}}

\put(2,20){\makebox(17,0){$\nu$}}
\put(6,10){\makebox(13,0){$\equiv$}}
\put(26,8){\makebox(-5,0){$p$}} \put(6,6){\makebox(-5,2){$1_U$}}
\end{picture}
\end{center}

\noindent In particular we use the term {\it global section} only
when $U=I$.

Given the category ${\cal A}$, one can form a new category ${\cal
A}^{op}$, called {\it dual} category of ${\cal A}$, by taking the
same objects but reversing the directions of all the arrows and
the order of all compositions. $Ens^{{\cal A}^{op}}$ or
$\widehat{{\cal A}}$ where ${\cal A}$ is a small category is the
category whose objects are functors $F: {\cal A}^{op} \rightarrow
Ens$ (also called {\it presheaves}) and whose arrows are natural
transformations between presheaves. $\widehat{{\cal A}}$ is a
topos, i.e. has terminal object, pullbacks, exponentiation and
subobject classifier. The terminal object in $\widehat{{\cal A}}$
is the functor ${\bf 1}: {\cal A}^{op} \rightarrow Ens$ such that
${\bf 1}(A) = \{*\}$ (the singleton) for each $A\in {\cal A}$ and
for each arrow $f$, ${\bf 1}(f) = \bf 1_{\{*\}}$. For any presheaf
$F:{\cal A}^{op} \rightarrow Ens$, the unique arrow $F \rightarrow
{\bf 1}$ is the natural transformation whose components are the
unique functions $F(A)\rightarrow {\{*\}}$ for each object $A$ in
${\cal A}$. Pullbacks, limits and colimits are defined
componentwise.

A {\it local section} of a presheaf $F: {\cal A}^{op} \rightarrow
Ens $ is a natural transformation $\tau: U \rightarrow F $ such
that $U$ is a subfunctor of the presheaf ${\bf 1}$. We only refer
to {\it global sections} in case that $U = {\bf 1}$.

\section{The question of valuation}

Let ${\mathcal H}$ be the Hilbert space associated to the physical
system and $L({\mathcal H})$ be the set of closed subspaces on
${\mathcal H}$. If we consider the set of these subspaces ordered
by inclusion, then $L({\mathcal H})$ is a complete orthomodular
lattice \cite{MM}. It is well known that each self-adjoint
operator $\bf A$ has associated a Boolean sublattice $W_A$ of
$L({\mathcal H})$. More precisely, $W_A$ is the Boolean algebra of
projectors ${\bf P}_i$ of the spectral decomposition ${\bf
A}=\sum_{i} a_i {\bf P}_i$. We will refer to $W_A$ as the spectral
algebra of the operator $\bf A$. Any proposition about the system
is represented by an element of $L({\mathcal H})$ which is the
algebra of quantum logic introduced by G. Birkhoff and J. von
Neumann \cite{byvn}.

Assigning values to a physical quantity {\it A} is equivalent to
establishing a Boolean homomorphism $v: W_A \rightarrow {\bf 2}$
\cite{isham1}, being ${\bf 2}$ the two elements Boolean algebra.
So it is natural to consider the following definition:

\begin{definition}

Let $(W_i)_{i\in I}$ be the family of Boolean sublattices of
$L({\mathcal H})$. A global valuation over $L({\mathcal H})$ is a
family of Boolean homomorphisms $(v_i: W_i \rightarrow {\bf
2})_{i\in I}$ such that $v_i\mid W_i \cap W_j = v_j\mid W_i \cap
W_j$ for each $i,j \in I$

\end{definition}

This global valuation would give the values of all magnitudes at
the same time maintaining a {\bf compatibility condition} in the
sense that whenever two magnitudes shear one or more projectors,
the values assigned to those projectors are the same from every
context.

But KS theorem assures that we cannot assign real numbers
pertaining to their spectra to operators ${\bf A}$ in such a way
to satisfy the functional composition principle (FUNC) which is
the expression of the ``natural'' requirement mentioned by Dirac
that, for any operator ${\bf A}$ representing a dynamical variable
and any real-valued function $f({\bf A})$, the value of $f({\bf
A})$ is the corresponding function of the value of ${\bf A}$. This
is a very restrictive constrain because it does not allow to
assign values to all possible physical quantities or to assign
true-false as truth values to all propositions about the system,
nor even non-contextual partial ones. KS theorem means that, if we
demand a valuation to satisfy FUNC, then it is forbidden to define
it in a non-contextual fashion for subsets of quantities
represented by commuting operators. In the algebraic terms of the
previous definition, KS theorem reads:

\begin{theo}\label{CS2}
If $\mathcal{H}$ is a Hilbert space such that $dim({\cal H}) > 2$,
then a global valuation over $L({\mathcal H})$ is not possible.
\qed
\\

\end{theo}

Of course contextual valuations allow us to refer to different
sets of actual properties of the system which define its state in
each case. Algebraically, a {\it contextual valuation} is a
Boolean valuation over one chosen spectral algebra. In classical
particle mechanics it is possible to define a Boolean valuation of
all propositions, that is to say, it is possible to give a value
to all the properties in such a way of satisfying FUNC. This
possibility is lost in the quantum case. And it is not a matter of
interpretation, it is the underlying mathematical structure that
enables this possibility for classical mechanics and forbids it in
the quantum case. The impossibility to assign values to the
properties at the same time satisfying FUNC is a weighty obstacle
for almost any interpretation of the formalism as something more
than a mere instrument.

\section{Spectral sheaf}
Being $L({\mathcal H})$ the lattice of closed subspaces of the
Hilbert space ${\mathcal H}$, we consider the family ${\cal W}$ of
all Boolean subalgebras of the lattice $L({\mathcal H})$ ordered
by inclusion and the topological space $\langle {\cal W}, {\cal
W}^+ \rangle$.  On the set $$E = \{(W,f): W\in {\cal W},\ f:W
\rightarrow  {\bf 2}\ \mbox{{\it is a Boolean homomorphism}}  \}
$$

\noindent we define a partial ordering given as

$$(W_1,f_1) \leq (W_2,f_2) \hspace{0.2cm} \Longleftrightarrow
\hspace{0.2cm} W_1 \subseteq W_2 \hspace{0.2cm} {\it and }
\hspace{0.2cm} f_1 = f_2 \mid W_1$$

\noindent Thus we consider the topological space $\langle E, E^+
\rangle$ whose canonical base is given by the principal decreasing
sets $((W,f)] = \{(G ,f\mid G): G \subseteq W \}$. By simplicity
$((W,f)]$ is noted as $(W,f]$.

\begin{prop}\label{SPECTRAL SHEAF}
The map $p:E \rightarrow {\cal W}$ such that $(W,f) \mapsto W$ is
a sheaf over ${\cal W}$.
\end{prop}

\begin{proof}
Let $(W,f) \in E$. If we consider the open set $(W,f]$ in $E$,
then $p((W,f]) = (W]$ resulting $p((W,f])$ an open set in ${\cal
W}$. If we denote by $p'$ the restriction $p\mid (W,f]$, then from
the definition of $p$ it is clear that $p': (W,f]\rightarrow (W]$
is a bijective map that preserve order inclusion. Thus $p'$ is a
continuous map. Finally $p$ is a local homeomorphism. \qed \\
\end{proof}

\noindent We refer to the sheaf $p:E \rightarrow {\cal W}$ as the
{\it spectral sheaf}.

\begin{prop}\label{LOCAL SECTION}
Let $\nu: U \rightarrow E$ be a local section of the spectral
sheaf $p$. Then for each $W\in U$ we have:

\begin{enumerate}
\item
$\nu(W) = (W,f)$ for some Boolean homomorphism $f:W \rightarrow
{\bf 2}$,

\item
if $W_0 \subseteq W$, then $\nu(W_0) = (W_0, f\mid W_0)$.

\end{enumerate}

\end{prop}

\begin{proof}
Since $\nu$ is a local section we consider the following
commutative diagram:

\begin{center}
\unitlength=1mm
\begin{picture}(60,20)(0,0)
\put(8,16){\vector(3,0){5}} \put(20,10){\vector(0,-2){5}}
\put(4,12){\vector(1,-1){10}}

\put(2,16){\makebox(0,0){$U$}} \put(20,16){\makebox(0,0){$E$}}
\put(20,0){\makebox(0,0){$U$}}

\put(2,20){\makebox(17,0){$\nu$}}
\put(6,10){\makebox(13,0){$\equiv$}}
\put(26,8){\makebox(-5,0){$p$}} \put(6,6){\makebox(-5,2){$1_U$}}
\end{picture}
\end{center}

\noindent 1) It follows as an immediate consequence of the
commutativity of the diagram.

\noindent 2) Since $\nu$ is continuous, $\nu ^{-1}((W,f])$ is an
open set in ${\cal W}$ (i.e. a decreasing set). Consequently $W_0
\in \nu ^{-1}((W,f])$ since $W_0 \subseteq W$ and $W \in \nu
^{-1}((W,f])$. Thus $\nu(W_0) \in (W,f]$ resulting $\nu(W_0) =
(W_0, f\mid W_0)$. \qed \\
\end{proof}

From the physical perspective, we may say that the spectral sheaf
takes into account the whole set of possible ways of assigning
truth values to the propositions associated with the projectors of
the spectral decomposition ${\bf A} = \sum_{i} a_{i} {\bf P}_i$.
The continuity of a local section of $p$ guarantees that the truth
value of a proposition is maintained when considering the
inclusion of subalgebras. In this way, the {\bf compatibility
condition} of the Boolean valuation with respect of intersection
of pairs of Boolean sublattices of $L({\mathcal H})$ is
maintained.\\

A global section $\tau: {\cal W} \rightarrow E $ of $p$ is
interpreted as follows: the map assigns to every $W \in {\cal W}$
a fixed Boolean valuation $\tau_w:W \rightarrow {\bf 2}$ obviously
satisfying the compatibility condition. So KS theorem in terms of
the spectral sheaf reads:

\begin{theo}\label{CS}
If $\mathcal{H}$ is a Hilbert space such that $dim({\cal H}) > 2$
then the spectral sheaf $p$ has no global sections. \qed \\

\end{theo}

We may build a {\it contextual valuation} in terms of a local
section as follows:\\

\noindent Let {\it A} be a physical magnitude with known value,
i.e. we have been able to establish a Boolean valuation $f:W_A
\rightarrow 2$. It is not very hard to see that the assignment\\

$\nu: (W] \rightarrow E$ such that for each $W_i\in (W]$,
$\nu(W_i) = (W_i, f\mid W_i)$\\

\noindent is a local section of $p$.\\

To extend contextual valuations we turn now to consider local
sections. To do this we introduce the following definition:

\begin{definition}

Let $\nu$ be a local section of $p$ and $W_A$ the spectral algebra
associated to the operator $\bf A$. Then an extended valuation
over $A$ is given by the set

$$\bar{\nu}(A) = \{W_B \in dom(\nu) : W_B\subseteq W_A \}$$

\end{definition}

Given the previous definition, it is easy to prove the following
proposition:

\begin{prop}\label{DECAT}
If $\nu$ is a local section of $p$ and $W_A$ the spectral algebra
associated to the operator $\bf A$, then:

\begin{enumerate}

\item

$\bar{\nu} (A)$ is a decreasing set,

\item

if $W_A\in U$ then $\bar{\nu}(A) = (W_A]$\qed\\

\end{enumerate}

\end{prop}

We can start from the spectral sheaf to build a representation as
presheaf such that local sections of the former are identifiable
to local sections of the later. When considering the family ${\cal
W}$ ordered by inclusion, ${\cal W}$ can be regarded as a small
category. Thus we can take the topos presheaf $\widehat{{\cal
W}}$. Denoting $E_W$ the fiber of the spectral sheaf $p$ over $W$
for each $W \in \cal{W}$, we consider the following presheaf:

$$D:{\cal W}^{op} \rightarrow Ens$$

\noindent such that

\begin{itemize}

\item
$D(W)=E_W$ for each $W\in Ob({\cal W})$.

\item
if $i:W_1 \subseteq W_2$ lies in $Ar({\cal W})$, then $D_i: D(W_2)
\rightarrow D(W_1)$ is such that $D_i(g) = g\mid W_1$.

\end{itemize}

It is clear that the presheaf acting over arrows satisfies the
compatibility condition. Denoting $Sec_p$ and $Sec_D$ the sets of
(global and local) sections of $p$ and $D$ respectively, we can
establish the following proposition:

\begin{prop}

$$Sec_p \simeq Sec_D$$

\end{prop}

\begin{proof}
Let $\nu :U\rightarrow E \in Sec_p$ and consider the presheaf
$\tilde{U}:{\cal W}^{op} \rightarrow Ens$ whose action over
$Ob({\cal W})$ is given by

$$ \tilde{U}(W) = \cases {\{*\}, & if $W\in U$  \cr \emptyset , &
$otherwise$\cr} $$

\noindent and whose action over arrows is immediate. It is clear
that $\tilde{U}$ is a subfunctor of the presheaf ${\bf 1}$. From
$U$ we construct a natural transformation

\begin{center}

\unitlength=1mm
\begin{picture}(60,20)(0,0)

\put(-3,13){\vector(3,0){12}}

\put(-3,3){\vector(3,0){12}}

\put(2,6){\vector(0,2){4}}

\put(2,16){\makebox(0,0){$D$}}

\put(14,8){\makebox(0,0){$Ens$}}

\put(-8,8){\makebox(0,0){${\cal W}^{op}$}}

\put(2,0){\makebox(0,0){$\tilde{U}$}}

\put(2,8){\makebox(-5,0){$\tau_\nu$}}

\end{picture}

\end{center}

\noindent such that, for each $W\in{\cal W}$, $\tau_\nu (U(W))=\nu
(W)$. Thus we have a map $Sec_p \rightarrow Sec_D$ given by $\nu
\mapsto \tau_\nu$. It is not hard to see that this is an injective
map. To see surjectivity, we consider a section of the presheaf
$D$, namely $\tau :\tilde{U}\rightarrow D$, and we prove that
there exists a section $\nu$ of the spectral sheaf such that $\nu
\mapsto \tau_\nu =\tau$. Let $U=\{ W\in {\cal W}: \tilde{U}(W) =\{
*\}\} $. If $W \in U$ and $W_o \subseteq W$, then $\tilde{U}(W_o
)=\{ *\}$ since $\tilde{U}$ is a contravariant subfunctor of {\bf
1}. Thus, $W_o$ lies in $U$, resulting $U$ a decreasing set (i.e.
an open set) in ${\cal W}$. Now we consider the map $\nu :U
\rightarrow E$ such that, for each $W\in U$, $\nu (W) = (W,\ \tau
(\tilde{U} (W)))$. It is clear that the following diagram is
commutative

\begin{center}
\unitlength=1mm
\begin{picture}(60,20)(0,0)
\put(8,16){\vector(3,0){5}} \put(20,10){\vector(0,-2){5}}
\put(4,12){\vector(1,-1){10}}

\put(2,16){\makebox(0,0){$U$}} \put(20,16){\makebox(0,0){$E$}}
\put(20,0){\makebox(0,0){$U$}}

\put(2,20){\makebox(17,0){$\nu$}}
\put(6,10){\makebox(13,0){$\equiv$}}
\put(26,8){\makebox(-5,0){$p$}} \put(6,6){\makebox(-5,2){$1_U$}}
\end{picture}
\end{center}

Now we prove that $\nu$ is a continuous map. Let $(W_1, \ f]$ be
an open set of the canonical base of $E$,  $\nu^{-1}((W_1, \
f])=\{W\in {\cal W}:\nu (W)\in (W_1, \ f]\}$ and we assume that
$\nu^{-1}((W_1, \ f])$ is not the empty set. Let $W\in
\nu^{-1}((W_1, \ f]) $ and $W_x \subseteq W$. Since $\tau$ is a
natural transformation, it follows that $\tau
(\tilde{U}(W_x))=\tau (\tilde{U}(W))|W_x$. Since $\nu (W)\in (W_1,
\ f]$, it is clear that $\nu (W)=(W, \ f|W)$, resulting that
$\tilde{U}(W_x)=f|W_x$ and $W_x \in \nu^{-1}((W_1, \ f])$. This
proves the continuity of the map. It is not very hard to see that
$\tau =\tau_\nu$, thus surjectivity is proved. \qed\\

\end{proof}

\begin{rema}

{\rm The presheaf $D$ from the spectral sheaf is the dual spectral
presheaf defined in Ref. \cite{isham1}}

\end{rema}

Taking into account the last Proposition, we can write KS theorem
in terms of presheaves from the spectral sheaf:

\begin{theo}\label{CS}
If $\mathcal{H}$ is a Hilbert space such that $dim({\cal H}) > 2$
then the dual spectral presheaf $D$ has no global sections. \qed
\\

\end{theo}

\noindent Possible obstructions to the construction of global
sections for the case of finite dimensional $\mathcal{H}$ are
shown in \cite{hamilton9912018}\\

On the other hand, in terms of a local section $\nu: U\rightarrow
D$ of $D$, extended contextual valuations over an operator $A$ may
be defined as

$$\bar{\nu}(A) = \{W_B\subseteq W_A: U(W_B)=\{*\} \}$$\\

Valuations are deeply connected to the election of particular
local sections of the spectral sheaf. So we see here once more
that we cannot speak of the value of a physical magnitude without
specifying this election, that clearly means the election of a
particular context. This is in agreement with the statement that
contextuality is ``endemic'' in any attempt to ascribe properties
to quantities in quantum theories \cite{isham1}

\section{Contextual logic}

We know that if $\cal{W}$ is the family of Boolean subalgebras of
$L({\mathcal H})$, to take a local section $\nu$ of the spectral
sheaf means an assignment of Boolean valuations to algebras in the
proper sub-family $Dom(\nu)$ maintaining the compatibility
condition. Now an interesting question is to ask what $\nu$ can
``tell us'' about $W$ when $W\not\in Dom(\nu)$. Let us state more
accurately this expression to precise our aim in the search of a
{\it contextual logic}.

\begin{definition}

Let $\nu$ be a local section of the spectral sheaf. If $W_B \in
Dom(\nu)$ and $W_B \subseteq W_A$ then we will say that $W_B$ has
{\it Boolean information} about $W_A$.
\end{definition}

Clearly this means that, in a given state of the system, the
complete knowledge of the spectral decomposition of $\bf{B}$ lets
us know the eigenvalue of one or more projectors in the spectral
decomposition of $\bf A$. {\it Contextual logic} allows some kind
of ``paste'' among Boolean sublattices of $L({\mathcal H})$ and so
among of CSCOs. A valuation in terms of decreasing sets maintains
it ``downstream'' with respect of subalgebras, i.e. when the
valuation of a subalgebra is given, all its subalgebras are
automatically valuated. This makes possible to have Boolean
information of different contexts from the one chosen in the
following sense: once fixed a local section $\nu$, if $W_B \in
Dom(\nu)$ and $W_A \not\in Dom(\nu)$ then, to witness $W_A$ from
$W_B$ refers to the Boolean information that $W_B \cap W_A$ has
about $W_A$.\\

We will now construct a propositional language $Self$ for
contextual logic whose atomic formulas refer to the physical
magnitudes represented for bounded self-adjoint operators with
discrete spectra. Intuitively we can consider the set of atomic
formulas ${\cal P}$ as $${\cal P} =\{A: A\hspace{0.2cm} {\it
bounded \ self-adjoint \ operator} \}$$ \noindent Then, this
language is conformed as follows:

$$Self = \langle {\cal P}, \lor, \land, \rightarrow, \neg
\rangle$$

\noindent  and it is clear that the formulas may be obtained in
the usual way.\\

We will now appeal to the use of Kripke models built starting from
any local section of the spectral sheaf $p$ because it allows to
naturally adapt the idea of Boolean knowledge. Thus the obtained
valuation will result in an extended contextual valuation.

\begin{definition}
{\rm We consider the poset $\langle {\cal W}, \subseteq \rangle $
as a $\it frame$ for the Kripke model for $Self$. Let $\nu$ be a
local section of $p$. Thus we define the Kripke model  ${\cal M} =
\langle {\cal W}, \bar \nu \rangle$ with the following forcing:

\begin{enumerate}

\item

${\cal M}\mid\models_W A$ \hspace{0.2cm} iff \hspace{0.2cm} ${\cal
W}\in \bar{\nu}(\bf{A})$\hspace{0.2cm} with $A \in {\cal P} $

\item

${\cal M}\mid\models_W \alpha \lor \beta$ \hspace{0.2cm} iff
\hspace{0.2cm} ${\cal M}\mid\models_W \alpha$ or ${\cal
M}\mid\models_W \beta$

\item

${\cal M}\mid\models_W \alpha \land \beta$ \hspace{0.2cm} iff
\hspace{0.2cm} ${\cal M}\mid\models_W \alpha$ and ${\cal
M}\mid\models_W \beta$

\item

${\cal M}\mid\models_W \alpha \rightarrow \beta$ \hspace{0.2cm}
iff \hspace{0.2cm} $\forall B\subseteq W $, if ${\cal
M}\mid\models_B \alpha$ then ${\cal M}\mid\models_B \beta$

\item

${\cal M}\mid\models_W \neg \alpha $ \hspace{0.2cm} iff
\hspace{0.2cm} $\forall B\subseteq W $ ${\cal M}\mid\not\models_B
\alpha$

\end{enumerate}

}

\end{definition}

Given this forcing we can accurately define the idea of extended
contextual valuation over $Self$.

\begin{definition}
Given a local section $\nu$ over $p$, an extended contextual
valuation is the map $\bar \nu : Self \rightarrow {\cal W}^{+}$
defined as

$$\bar \nu(\alpha) = \{W: {\cal M}\mid\models_W \alpha \}$$
\end{definition}

\noindent Taking into account that ${\cal W}^{+}$ is a topological
space it is not very hard to see that $\bar \nu(\alpha)$ is an
open set of ${\cal W}$. Now we can to establish the following
proposition:

\begin{prop}\label{BORDER}

Let $\alpha$ be a formula in $Self$ and consider the Kripke model
${\cal M} = \langle {\cal W}, \nu  \rangle$. Then:

\begin{enumerate}

\item
${\cal M}\mid\models_W \neg \alpha $ \hspace{0.2cm} iff
\hspace{0.2cm} $W\in (C \bar \nu(\alpha))^\circ$

\item
${\cal M}\mid\not\models_W \alpha$ \hspace{0.2cm} and
\hspace{0.2cm} ${\cal M}\mid\not\models_W \neg \alpha$
\hspace{0.3cm} iff \hspace{0.3cm} $W\in \partial \bar \nu(\alpha)$

\end{enumerate}

\end{prop}

\begin {proof}
\noindent 1) If  ${\cal M}\mid\models_W \neg \alpha $,
\hspace{0.2cm} then \hspace{0.2cm} $\forall B\subseteq W $, ${\cal
M}\mid\not\models_B\alpha$ \hspace{0.2cm} and \hspace{0.2cm}
$\forall B\subseteq W $ $B \not \in \bar \nu(\alpha)$. Thus
$(W]\subseteq (C \bar \nu(\alpha))^\circ$ and $W\in (C \bar
\nu(\alpha))^\circ$. On the other hand, if $W \in (C \bar
\nu(\alpha))^\circ$, then there exists an open set $G$ in $\cal W$
such that $W \in G \subseteq C( \bar \nu(\alpha))$. Since $G$ is a
decreasing set, we have that $(W] \subseteq G \subseteq C( \bar
\nu(\alpha))$ and ${\cal M}\mid\models_W \neg \alpha $. \noindent
2)It follows from 1) and the fact that ${\cal W} = \bar
\nu(\alpha) \cup (C \bar \nu(\alpha))^\circ \cup
\partial \bar \nu(\alpha))$ \qed
\end {proof}

\begin{rema}

{\rm Following the usual interpretation of the Kripke model, the
frame represents all possible states of knowledge that are
preserved forward in time. In our case, the frame $\langle {\cal
W}, \subseteq \rangle $, ${\cal W}$ represents all states of
Boolean knowledge in the sense of all possible Boolean valuations
of spectral algebras and the usual notion of ``preserving
knowledge through time'' must be understood in terms of
$\subseteq$ as ``preserving valuations in spectral subalgebras''.
The forcing $K\mid\models_W \alpha $ is interpreted as {\it the
spectral algebra $W$ has Boolean knowledge about $\alpha$} i.e.
the complete Boolean valuation of $W$ is known and $W$ lies in the
decreasing set associated to the formula $\alpha$. By Proposition
\ref{DECAT}, to know the eigenvalue of {\bf A} is expressed in
terms of the forcing as ${\cal M}\mid\models_{(W_A]} A $.}

\end{rema}

Being ${\cal W}^{+}$ a topological space, it is a Heyting algebra
with meet and join operations the classical ones and implication
and negation defined as follows: \\

$S\rightarrow T = \{P\in {\cal W}: \forall X\subseteq
P,\hspace{0.2cm} if \hspace{0.2cm} X\in S \hspace{0.2cm} then
\hspace{0.2cm} X\in T \}$\\

$\neg S = \{P\in {\cal W}: \forall X\subseteq P,\hspace{0.2cm}
X\notin S \}$\\

\noindent Thus the extended contextual valuation is a Heyting
valuation of $Self$ from the Heyting algebra ${\cal W}^{+}$ such
that

\begin{enumerate}

\item
$\bar \nu(\alpha \lor \beta) = \bar \nu(\alpha) \cup \bar
\nu(\beta)$

\item
$\bar \nu(\alpha \land \beta) = \bar \nu(\alpha) \cap \bar
\nu(\beta)$

\item
$\bar \nu(\alpha \rightarrow \beta) = \bar \nu(\alpha) \rightarrow
\bar \nu(\beta)$

\item
$\bar \nu(\neg \alpha) = \neg \bar \nu(\alpha) $

\end{enumerate}

Taking into account the restrictions in the valuations imposed by
the KS theorem, it is not possible a Heyting valuation $v: Self
\rightarrow {\cal W}^{+}$ such that $v(A) = (A]$ for each atomic
formula $A$. So it is clear that contextual logic is an
intuitionistic logic in which not all of the Heyting valuations
are allowed.

\section{Conclusions}

Contextual logic is a formal language to deal with combinations of
propositions about physical properties of a quantum system that
are well defined in different contexts. These properties are
regarded from a fixed context, which guarantees the avoidance of
no-go theorems. This means that one can refer to contexts other
than the chosen one by building a Kripke model in which each
proposition is given a decreasing set as its truth value.

There are different formal languages on the orthomodular lattice
of closed subspaces of $\mathcal H$ (as orthologic or orthomodular
quantum logic), but these logics give rise to different problems
that lack an intuitive understanding, as the ``implication
problem'' (briefly, eight different connectives may represent the
material conditional, see \cite{dallachiara} ). On the contrary,
as contextual logic is an intuitionistic one -with restrictions on
the allowed valuations arising from the KS theorem- it has
``good'' properties as the distributive lattice structure and a
nice definition of the implication as a residue of the
conjunction. The price paid is being a contextual language. But
this is not a difficulty, it is a main feature of quantum
mechanics.

{\small \noindent Hector Freytes: \\ Escuela de Filosof\'{\i}a \\
Universidad Nacional de Rosario,\\ Entre R\'{\i}os 758, 2000,
Rosario, Argentina \\e-mail: hfreytes@dm.uba.ar}

{\small \noindent Graciela Domenech: \\ Instituto de
Astronom\'{\i}a y F\'{\i}sica del Espacio (IAFE)\\Ciudad
Universitaria\\1428 Buenos Aires - Argentina\\e-mail:
domenech@iafe.uba.ar}

\end{document}